\documentclass[reqno,10pt,centertags]{amsart}
\usepackage{amssymb,upref,graphicx,lastpage,cite}

\newcommand{\be}{\begin{equation}}
\newcommand{\ee}{\end{equation}}

\allowdisplaybreaks \numberwithin{equation}{section}

\usepackage{graphicx,lastpage}
\makeatletter
\def\@maketitle@hook
  {\hbox to \textwidth
    {\valign{\vss##\vss\cr
      \hbox{\includegraphics[width=30mm]{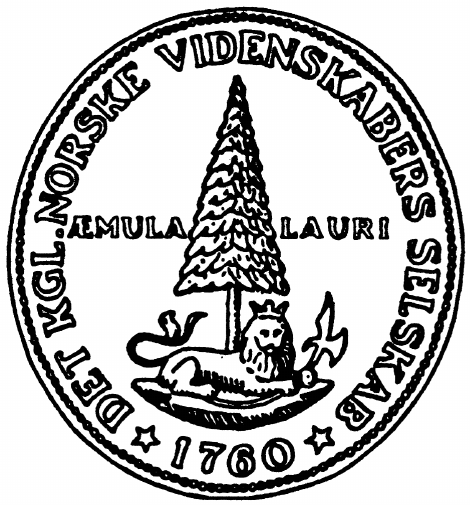}}\cr
      \noalign{\hfil}
      \hsize=0.7\hsize
      \vbox{\parindent0pt \raggedleft

         \medskip
         \sffamily
         Transactions of The\\ Royal Norwegian Society\\ of Sciences and
         Letters,\\ 
         (\textit{Trans.\ R.\ Norw.\  Soc.\  Sci.\ Lett.} 20xx(x) x--x)\par}\cr}}
  \vskip 42pt \relax}
\makeatother

\begin{document}

\title[]{Nonextensive statistical dynamics applied to wall turbulence}

\author[]{Simen \AA.\ Ellingsen}
\author[]{Per-\AA ge Krogstad}
\address{Department of Energy and Process Engineering, Norwegian University of Science and Technology, N-7491 Trondheim, Norway}

\dedicatory{Dedicated  to Johan S.\ H\o ye on his 70th birthday.}
\date{\today}

\begin{abstract}
  We apply a formalism of nonextensive statistical mechanics to experimental wall turbulence data, for the first time to our knowledge. Wind tunnel data for velocity differences a streamwise distance $r$ apart are compared to the prediction from theory as developed by Beck. The simplest theory, in which all free parameters are removed, is found to reproduce statistics for the wall-normal velocity component remarkably well, even for $r$ well beyond the corresponding integral scale, while the corresponding description of the streamwise velocity fluctuations is reasonable at separations below the integral scale. A least-squares 2-parameter fit is performed, and the dependence of the optimum parameter values on wall separation and $r$ is analysed. Both parameters are found to be approximately independent of wall-separation in the logarithmic sub-layer.
\end{abstract}

\maketitle

\section{Introduction}

Some time ago, Tsallis \cite{tsallis88,tsallis98} suggested a generalisation of statistical mechanics which has since attracted much attention. In this generalisation the standard Boltzmann--Gibbs entropy $S = -\sum_i p_i \ln p_i$ is replaced by
\be
  S_q = \frac1{q-1}\left(1-\sum_i p_i^q\right),
\ee
where $\{i\}$ is the set of possible microstates and $p_i$ their respective probabilities. $q$ is a real number. Boltzmann--Gibbs statistics are regained in the limit $q\to 1$ whereas $q>1$ represents statistics placing more emphasis on large but rare fluctuations. Tsallis statistics have been found to describe a broad variety of systems, including, among others, cold atoms in optical lattices \cite{lutz03,douglas06}, high-energy particle physics \cite{alberico00}, quantum scattering \cite{ion99}, granular flow \cite{sattin03,arevalo07} and many other topics within physics, chemistry, biology, computer science and economics. For reviews of foundations and applications of nonextensive statistical mechanics, see e.g., \cite{abe01,tsallis09}.

Probably the most researched application of non-extensive Tsallis statistics in physics has been in the field of turbulent flows, where a sizeable literature exists. For the case of fully developed three-dimensional turbulence, some analyses were made reporting values $0<q<1$ \cite{arimitsu00,arimitsu00b}, while a very elegant theory based on $q>1$ was developed by Beck \cite{beck00} and applied to different experimental data sets \cite{beck01,beck01b,beck01c,beck01d,beck02,beck02b}. In the following we will apply Beck's theory to the case of wall turbulence, a case not previously studied in this manner to the best of the authors' knowledge.

In section \ref{sec:pdf} Beck's theory for the probability density function (PDF) of velocity differences in fully developed turbulence is briefly recounted, with consideration made for the particular case of wall turbulence. Section \ref{sec:exp} contains the comparison with experimental wall turbulence data from a wind tunnel, with accompanying discussion, before conclusions and final remarks.

\section{Probability density function for turbulent velocity} \label{sec:pdf}

We consider the difference between simultaneous velocities in two points separated by a distance $r$ in the streamwise direction. Experimentally this is realised via Taylor's frozen eddy hypothesis whereby turbulent structures are presumed to be transported with the mean flow. Beck's theory \cite{beck00} was designed for differences in the radial component of velocity with respect to the two points, i.e., the streamwise component in our case, yet we shall apply it also to differences in the wall-normal velocity component in the following. There seems no particular reason why only the radial component should be used, and indeed we find that differences in the wall-normal component are better described by the theory than is the case for the streamwise.

The PDF is designed in close analogy to standard equilibrium statistical mechanics, by extremising the Tsallis entropy $S_q$, whereby a state with energy $\epsilon_i$ obtains the probability 
\begin{subequations}
\begin{align}
  p_i =& Z_q^{-1}[1-(1-q)\beta\epsilon_i]^{-1/(q-1)}; \\
  Z_q =& \sum_i [1-(1-q)\beta\epsilon_i]^{-1/(q-1)}.
\end{align}
\end{subequations}
Here $\beta$ plays the role of an inverse ``temperature''. In the limit $q\to 1$, $p_i\propto \exp(-\beta \epsilon_i)$ is obtained as for the Boltzmann--Gibbs case. 

For a stationary flow we may split the absolute fluid velocity field into a mean and a fluctuating term, $\mathbf{v}_\text{abs} = \langle\mathbf{v}\rangle+ \mathbf{v}$. For succinctness, let $u$ denote the difference between two velocity components a streamwise distance $r$ apart, nondimensionalised by its root-mean-square (rms) turbulent velocity component. The streamwise direction we denote $\hat{\mathbf{x}}$ and the wall-normal $\hat{\mathbf{z}}$, and corresponding velocity components $v_x$ and $v_z$, hence $u$ is either $u=[v_x(\mathbf{x}+r\hat{\mathbf{x}})-v_x(\mathbf{x})]/\mathrm{rms}[ v_x(\mathbf{x}+r\hat{\mathbf{x}})-v_x(\mathbf{x})]$ or $u=[v_z(\mathbf{x}+r\hat{\mathbf{x}})-v_z(\mathbf{x})]/\mathrm{rms}[ v_z(\mathbf{x}+r\hat{\mathbf{x}})-v_z(\mathbf{x})]$ ($\mathrm{rms}$ denotes the root-mean-square) at some position $\mathbf{x}$. For our experimental data we may assume to a good approximation that mean quantities vary only in the $z$ direction (all means taken are sample averages, equal to time averages for practical purposes). Obviously, $u$ has unit variance and zero mean.

As an ``energy'' $\varepsilon(u)$, Beck suggests the form \cite{beck00}
\be
  \epsilon(u) = \frac1{2}u^2 - C\Bigl(u - \frac{1}{3}u^3\Bigr)
\ee
where $C$ is a small constant. The form of the quadratic term forming the main contribution is easily recognisable as the kinetic energy density. The asymmetric term is necessary to reproduce the skewness observed in experimentally measured PDFs. Motivation is provided by the observation that a large class of chaotic maps produce PDFs characterised by an ``energy'' of just this form up to corrections of order $C^2$ \cite{beck00}, and thus this seems to represent some level of universality. Naturally such a form for the skewness term can only hold for small $u$, lest the cubic term will eventually dominate as $u$ grows large. We therefore replace it with a Pad\'{e} approximant with the same asymptotics as $u\to 0$, whereby $u - u^3/3 \longrightarrow u/(1+u^2/3)$.

The PDF of $u$ may thus be written
\be
  p(u) = \frac1{Z_q}\Bigl[1+\frac{2(q-1)}{5-3q}\Bigl(\frac{u^2}{2}- \frac{Cu}{1+u^2/3}\Bigr)\Bigr]^{-\frac{1}{q-1}}
\ee
where $\beta=2/(5-3q)$ has been inserted to ensure $\langle u^2\rangle=1$ in the limit $C\to 0$ ($1\leq q<5/3$ is assumed). This PDF has a small non-zero expectation value of order $C$; this is rectified in later analysis by simply shifting the origin and comparing to the function $p(u-\langle u\rangle)$.

\subsection{Physical interpretation of parameter $q$}

An interpretation of the parameter $q$ is proposed in Ref.~\cite{beck00} and more easily seen if introducing another parameter $k=1/(q-1)$;
\be\label{pdf}
  p(u) = \frac1{Z_q}\Bigl[1+\frac{2}{2k-3}\Bigl(\frac{u^2}{2}- \frac{Cu}{1+u^2/3}\Bigr)\Bigr]^{-k}.
\ee
An understanding of $k$ is proposed by looking to the turbulence cascade.

Turbulence is produced in the form of large eddies by the shear resulting from the fluid passing the solid wall. The eddies interact, are stretched and break up into ever smaller eddies, until their size is of the order of the Kolmogorov microscale $\eta = (\nu^3/\varepsilon)^{1/4}$ ($\nu$: kinematic viscosity, $\varepsilon$: dissipation of turbulent kinetic energy). $\eta$ is the smallest turbulent lengthscale, at which energy is not transferred further but dissipated to heat. While somewhat simplistic, this picture of the turbulent cascade is a cornerstone in our understanding of turbulent processes.

As $r\to \infty$ the velocities in the two points become independent, and the distribution of $u$ is farily close to Gaussian, corresponding to $k\to \infty$. When $r \sim \eta$, on the other hand, the separation between the two points is as small as the smallest eddies, and the velocities must be strongly correlated, producing highly non-Gaussian PDFs, as produced for $k = \mathcal{O}(1)$. Beck suggests to interpret $k$ as simply the level of the turbulent cascade, where the eddy size is presumed to double for each step. The bold assumption is thus
\be\label{kfit}
  \frac{r}{\eta} = 2^{k-1}
\ee
omitting a possible proportionality constant of order unity. Knowing the complexity of all things turbulent, one would be forgiven for thinking such a relation, which eradicates one of only two free parameters of the model, too good to be true. Remarkably, however, good correspondence is found with different sets of experimental data \cite{anselmet84,castaing90,chabaud94}. 

\begin{figure}[htb]
  \includegraphics[width=.7\textwidth]{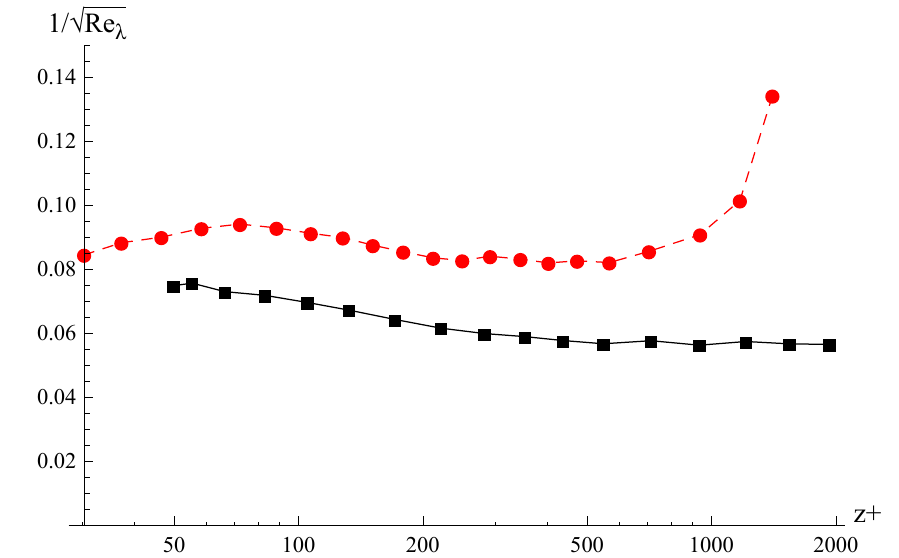}
  \caption{$(\mathrm{Re}_\lambda)^{-1/2}$ as a function of wall separation $z^+$ for the data sets $\mathrm{Re}_\theta=12600$ ({\scriptsize $\blacksquare$}) and $\mathrm{Re}_\theta=4350$ ({\LARGE$\bullet$}).}
\end{figure}

As for the skewness constant $C$, Beck argues it should scale with $\mathrm{Re}_\lambda^{-1/2}$, where $\mathrm{Re}_\lambda$ is the Reynolds number based on the Taylor microscale and the turbulent velocity $\sqrt{\langle \mathbf{u}^2\rangle}/3$. In Ref.~\cite{beck00} good agreement with the experimental data considered therein is found by simply using $C=\mathrm{Re}_\lambda^{-1/2}$, leaving no free parameters at all. For the case of wall turbulence, we find that this must be modified somewhat for concordance with experiment.

\section{Comparison with experimental data}\label{sec:exp}

\begin{figure}[htb]
  \includegraphics[width=.49\textwidth]{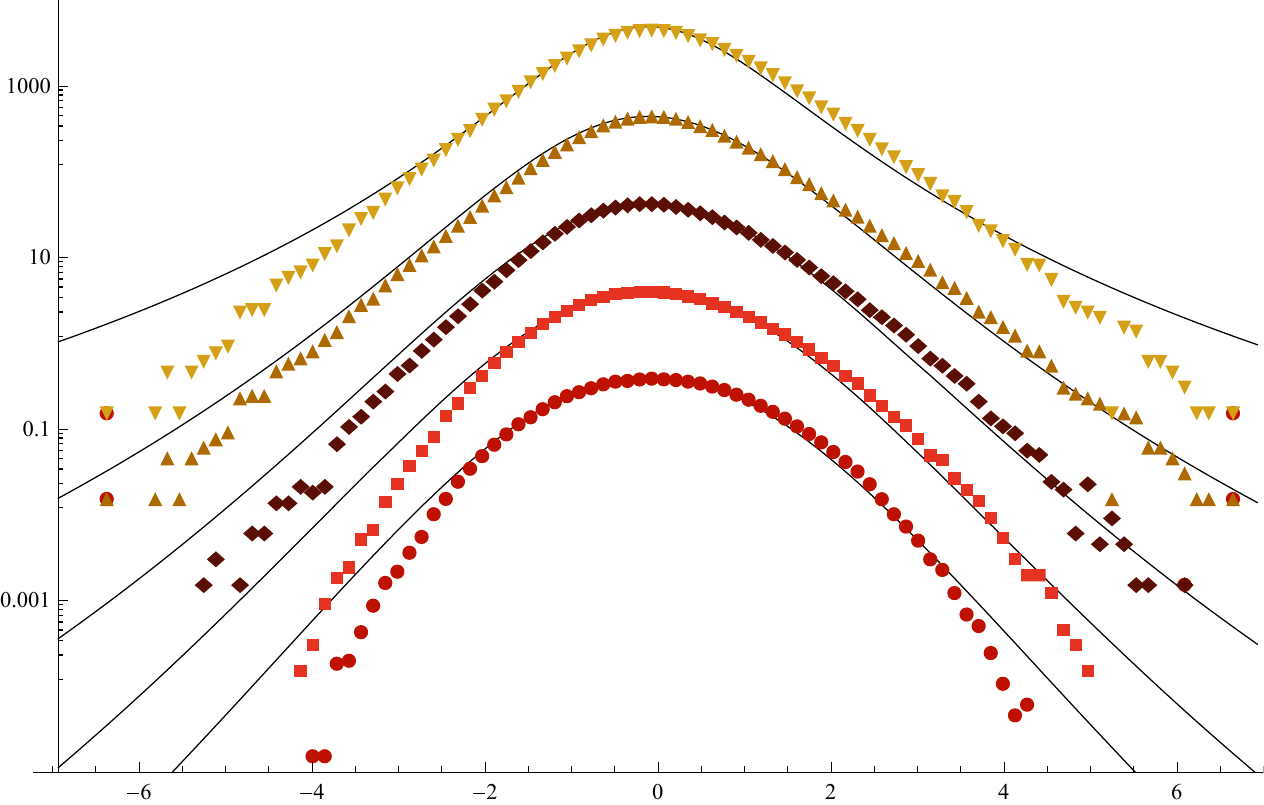}~
  \includegraphics[width=.49\textwidth]{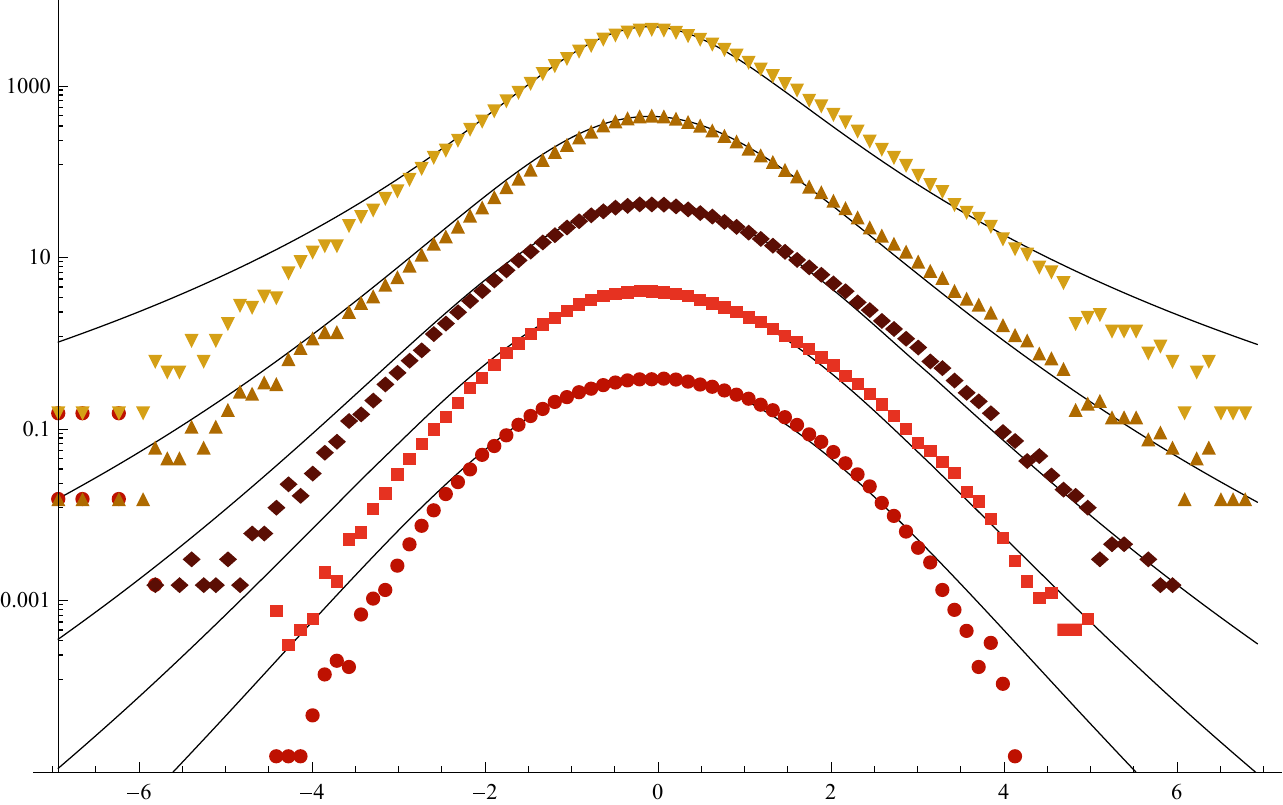} \\
  \includegraphics[width=.49\textwidth]{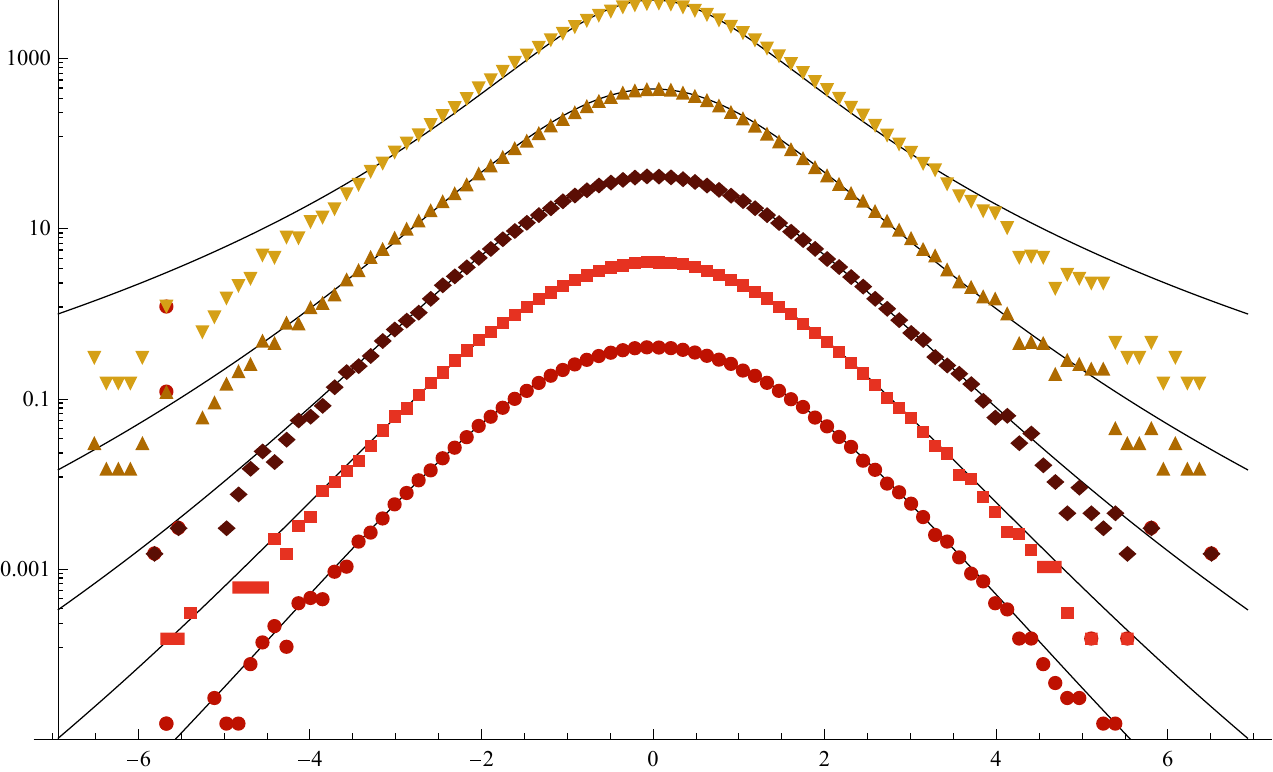}~
  \includegraphics[width=.49\textwidth]{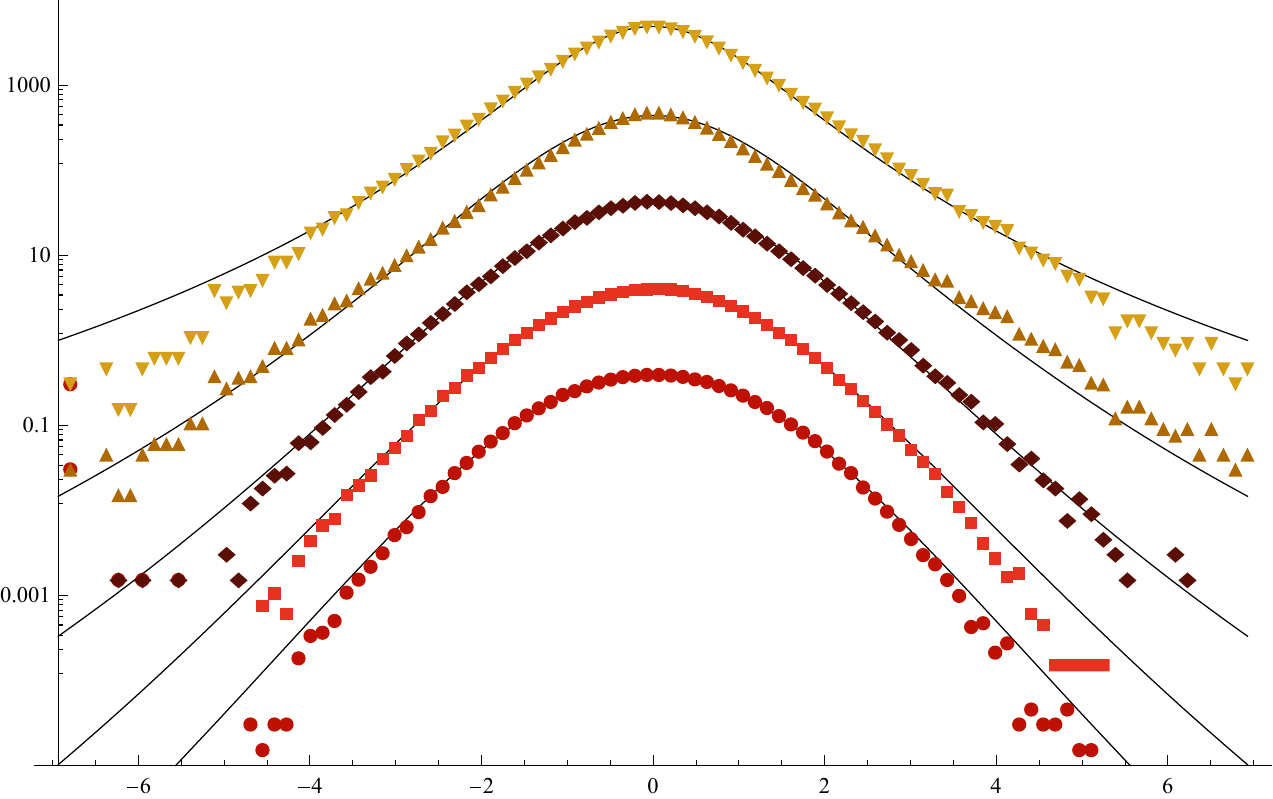}
  \caption{Comparison of experimentally measured PDF of velocity differences with simplest, parameter-free model, at $\mathrm{Re}_\theta=12600$ at $z^+=105$ (left) and $z^+=1211$ (right), for streamwise (top) and wall-normal (bottom) velocity components. In each panel graphs are, from top to bottom, $r/\eta \approx 4, 16, 64, 256, 1024$ (corresponding to $k=3,5,7,9,11$), separated by factors of $10$ for visibility.}
  \label{fig:naturals}
\end{figure}

Experimental data were obtained by hot-wire measurement in the wind tunnel at the Norwegian University of Science and Technology in Trondheim, Norway, as previously presented in Ref.~\cite{davidson06}. The primary data set used for comparison consisted of simultaneously measured streamwise and wall-normal velocity components at Reynolds number $\mathrm{Re}_\theta=12600$ (based on the momentum thickness  $\theta$). A second data set at $\mathrm{Re}_\theta=4350$ was used for comparison. External stream velocities were $24.9$ and $8.75$ m/s, respectively, and sampling frequency $30$ and $20$ kHz. Results in the present context are qualitatively similar for both sets, except in a few respects as indicated in the following.

The theoretical basis for Tsallis statistical mechanics in a turbulence context has been argued for the case of fully developed and statistically steady turbulence \cite{beck01}. In the case of wall turbulence this may be taken to mean positions sufficiently far downstream that the boundary layer has developed a self-similar (Reynolds number independent) logarithmic sub-layer. This is true for $\mathrm{Re}_\theta$ greater than a few thousand (for example, DeGraaff and Eaton find it to be true for $\mathrm{Re}_\theta\gtrsim 2000$ \cite{degraaff00}). There is thus reason to expect the statistics of both data sets to be captured at least to some extent by the simple theory.

A fundamental difference between wall turbulence and the turbulence data for which the model has previously been tested, is the introduction of an additional physical scale, the separation between measuring point and the wall, $z$.  
In principle, parameters $k$ (i.e., $q$) and $C$ could vary with $z$ in very general ways. The hypothesis that is suggested by the considerations of Ref.~\cite{beck00} is that in the regions of the boundary layer where self-similarity is present, the variation of the best values of $k$ and $C$ should be indirect, through the local variations of $\eta(z)$ and $\mathrm{Re}_\lambda(z)$ only. Our results are not inconsistent with this hypothesis, although this investigation does not amount to a systematic test.

As is customary, we use wall units to denote separations; the nondimensional wall separation is $z^+= z U_*/\nu$ ($U_*$ is the standard friction velocity). As rule of thumb, the mean velocity profile is well described by the log-law of the wall in the range $40\lesssim z^+ \lesssim 700$ \cite{davidson04}, which is called the logarithmic sub-layer or log-layer. Such a layer is characterised by $z^+\gg 1$ while at the same time $z \ll \delta$, $\delta$ being the boundary layer thickness; as mentioned such a range exists for sufficiently high Reynolds numbers. We shall consider separations roughly in this range, which is where reasonable fits can be found; indeed turbulence statistics either inside or outside this range have features which such a simple theory cannot hope to capture (strong influence from viscous shear in the inner region, and intermittent gusts from large coherent structures in the outer).

\subsection{Attempt at a parameter-free model}

Using experimental data at $\mathrm{Re}_\theta=12600$, we first wish to investigate how the model PDF (\ref{pdf}) fares when applied as in Ref.~\cite{beck00}, using assumptions to remove free parameters. The results are representative also for $\mathrm{Re}_\theta=4350$ for $z^+$ well inside the log-layer. For the streamwise ($v_x$) component we use the relation $C=-1/\sqrt{\mathrm{Re}_\lambda}$ while for wall-normal ($v_z$) we use $C=0$. The only fitting that has been done thus consists in a change of sign for the streamwise $C$ compared to \cite{beck00} to match the observed sign of the skewness, and an observation that there is much less skewness for the wall-normal component. As suggested in \cite{beck00} we use $k = \log_2(r/\eta)+1$, Eq.~(\ref{kfit}). 

As figure \ref{fig:naturals} shows, the model does a remarkably good job of fitting the data, considering its simplicity and that very little parameter fitting is involved. Overall the fit is better for the wall-normal component, despite the fact that the model was designed for and applied to streamwise components in the past. For both components the fit is fairly good for $k=3,5,7$, while deviations become noticeable for the streamwise component for $k=9$ and $11$. We shall argue in a moment that this is to be expected, while the true surprise is in fact that the fit for $v_z$ remains good up to $k=11$.

Clearly, the cascade interpretation of $k$ can only work for values of $r$ that are larger than the smallest eddies and smaller than the largest eddies in the flow. The largest dimensions of coherent structures is called the integral scales, and we must expect the cascade picture of $k$ to break down for $r\gtrsim L$ where $L$ is the relevant integral scale.
A much used way to define the streamwise integral scale in wall turbulence for the streamwise and wall-normal components is  
\be
  L_{xx}^{(x)} = \int_0^\infty\mathrm{d} r\frac{\langle v_x(0)v_x(r\hat{\mathbf{r}})\rangle}{\langle v_x^2(0)\rangle},~~ L_{zz}^{(x)} = \int_0^\infty\mathrm{d} r\frac{\langle v_z(0)v_z(r\hat{\mathbf{r}})\rangle}{\langle v_z^2(0)\rangle}.
\ee
At the two positions used in figure \ref{fig:naturals}, we use experimental data to evaluate these to $L_{xx}^{(x)}/\eta = 419, L_{zz}^{(x)}/\eta=77.4$ at $z^+=105$ and $L_{xx}^{(x)}/\eta = 391, L_{zz}^{(x)}/\eta=82.1$ at $z^+ = 1211$. This implies that the maximal $r$ for which the relation (\ref{kfit}) is expected to be reasonable, corresponds to about $k=9.6$ for streamwise and $k=7.3$ for wall-normal components. Beyond this range, the cascade picture is expected to fail. 

It is striking then, that while the streamwise component fails to be well approximated at $r\sim 2^8\eta$ as expected, the wall-normal component appears to follow the $k=\log_2(r/\eta)+1$ scaling reasonably closely all the way to $k=11$, corresponding to separations four orders of magnitude beyond the $zz$-integral scale and two orders beyond the $xx$-integral scale. 

Another gratifying observation is that an ``energy'' that goes like $u^2$ does a good job of reproducing observed PDF, while in the Taylor--Couette flow of Ref.~\cite{beck01b}, a more general exponent was found to match better: $2 \to 2(2-q)$, a phenomenon which is difficult to explain. With the wall-turbulence data, an exponent of $2$ is close to optimal, as would be expected from Newtonian mechanics.

\subsection{Best fit of two-parameter model}

\begin{figure}[htb]
  \includegraphics[width=.7\textwidth]{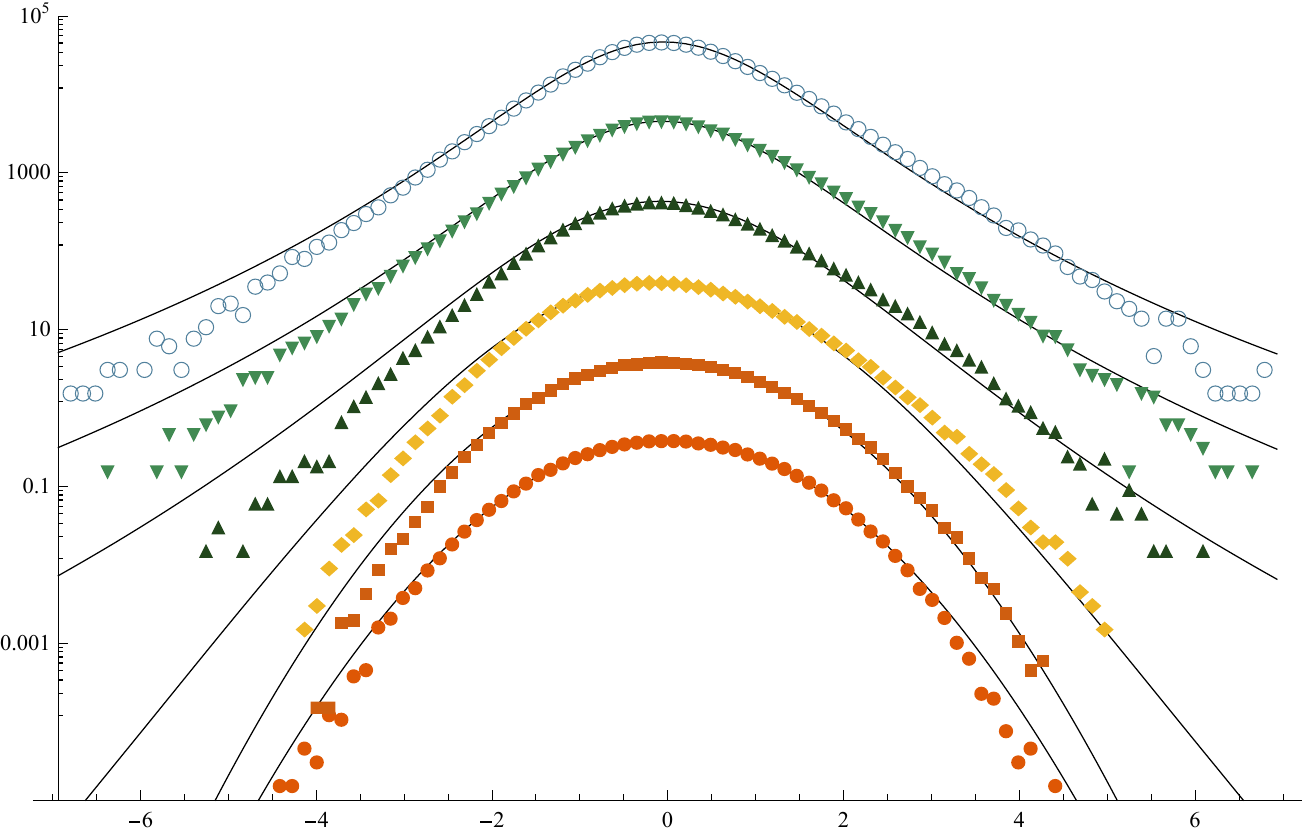} \\
  \includegraphics[width=.7\textwidth]{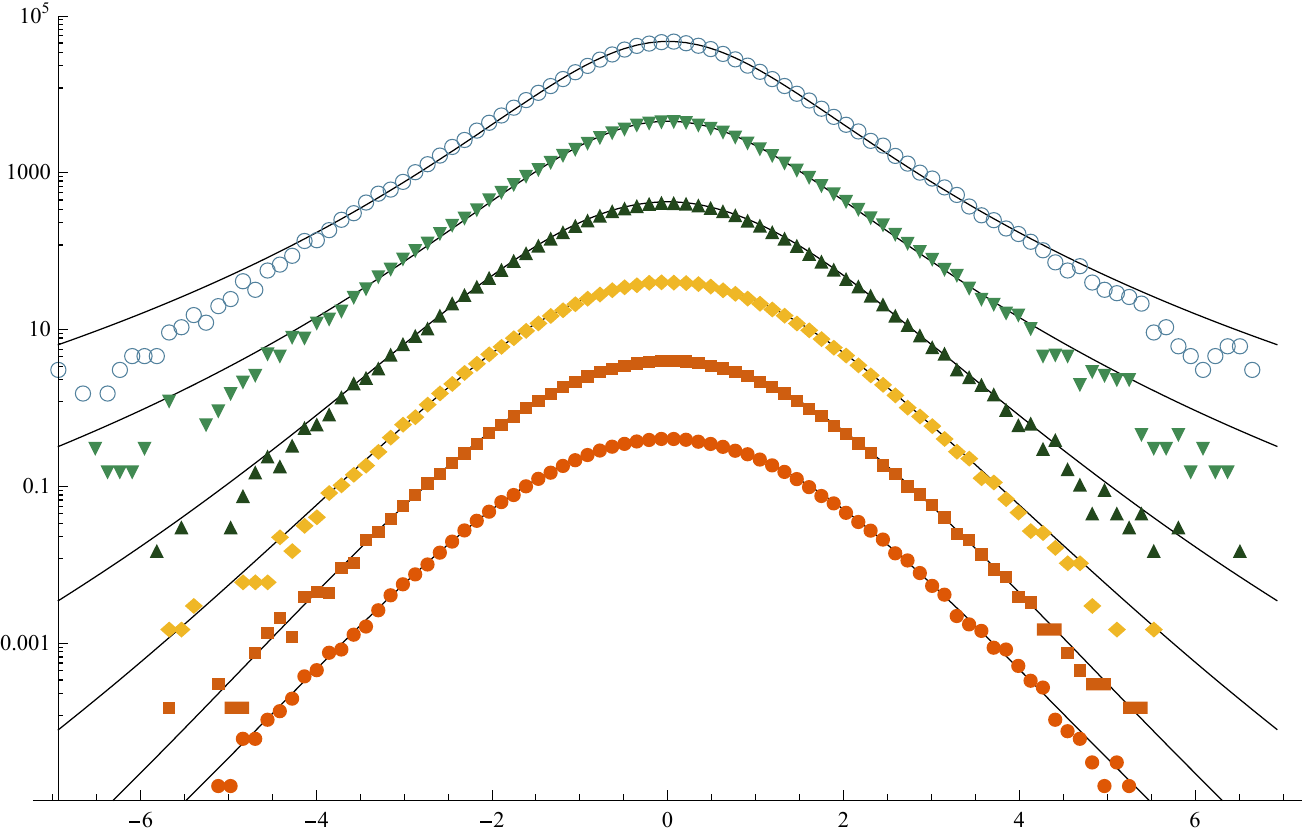} 
  \caption{Experimentally measured PDF of velocity differences and best 2-parameter fits at $\mathrm{Re}_\theta=12600$ at $z^+=105$, for streamwise (above) and wall-normal (below) velocity components. In each panel graphs are, from top to bottom, $r/\eta \approx 4, 16, 64, 256, 1024,4096$, separated by factors of $10$ for visibility.}
  \label{fig:bestfits}
\end{figure}

\begin{table}[htb]
  \begin{tabular}{c|cccccc}
    \hline
    $\log_2(r/\eta)+1$ & 3&5&7&9&11&13 \\
    \hline
    Best fits $v_x$, $k$ & 3.67 & 4.17 & 5.95 & 18.6 & 253 & 185\\
    Best fits $v_z$, $k$ & 3.42 & 4.11 & 6.92 & 9.49 & 12.6 & 11.9\\
    Best fits $v_x$, $C$ & -0.031 & -0.034 & -0.055 & -0.056 & -0.024 &-0.0029\\
    Best fits $v_z$, $C$ & 0.0028 & 0.0040 & 0.0021 & 0.0046 & 0.0015 & 0.0012\\
    \hline
  \end{tabular}
  \caption{Best fit values for $k$ and $C$ for graphs in figure \ref{fig:bestfits}.}
  \label{tab:kC}
\end{table}

We now go on to determine the values of $k$ and $C$ which yield the closest fit between the theoretical PDF (\ref{pdf}) and the experimental data. We use Mathematica's \texttt{FindFit} function to obtain least squares estimates for the two parameters for different values of separation $r/\eta$ and wall separation $z^+$. 

\begin{figure}[htb]
  \includegraphics[width=.49\textwidth]{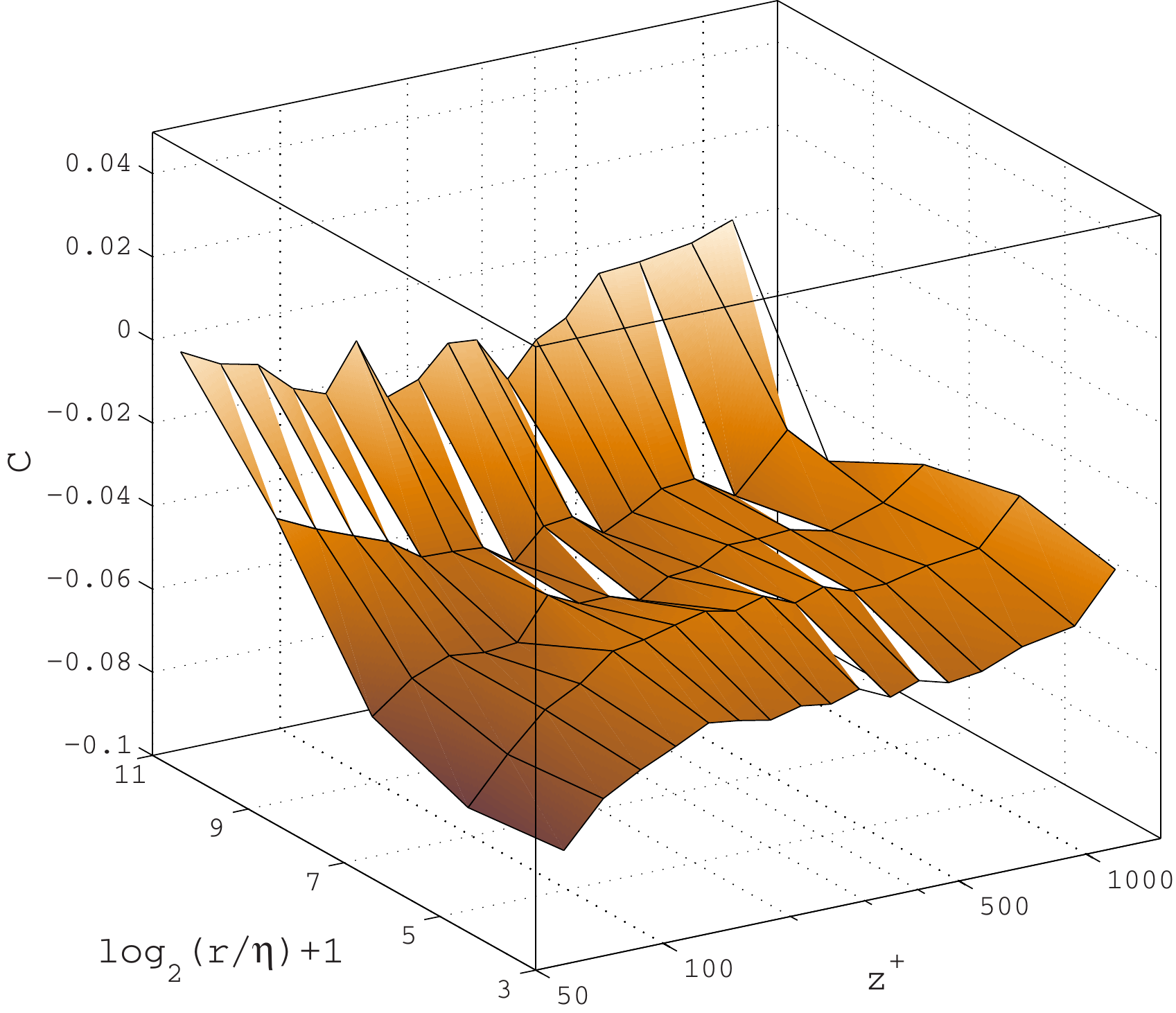}  \includegraphics[width=.49\textwidth]{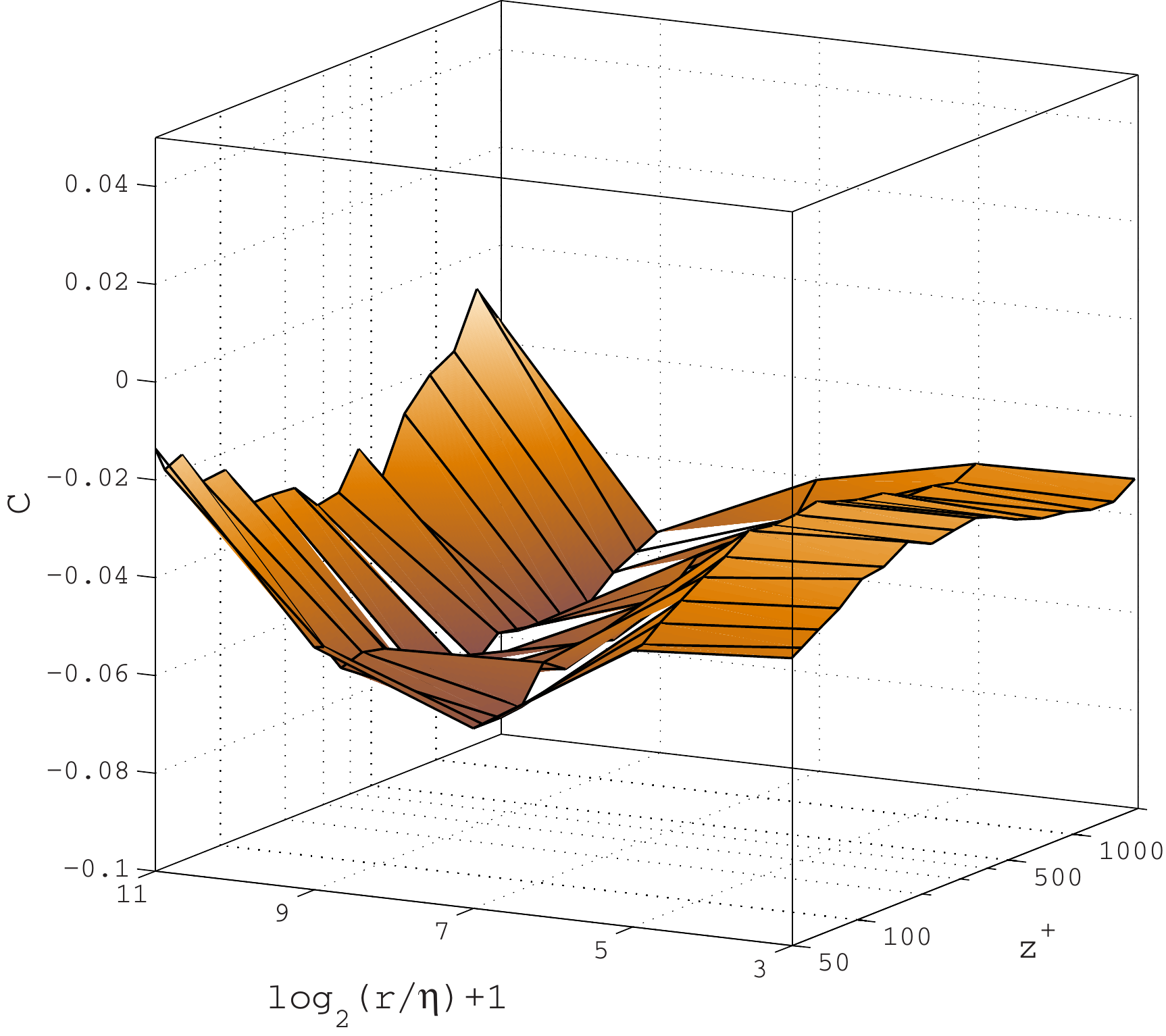} 
  \caption{Best fits of $C$ for the streamwise velocity component at $\mathrm{Re}_\theta=4350$ (left) and $12600$ (right).}
  \label{fig:Cfits}
\end{figure}

\begin{figure}[htb]
  \includegraphics[width=.49\textwidth]{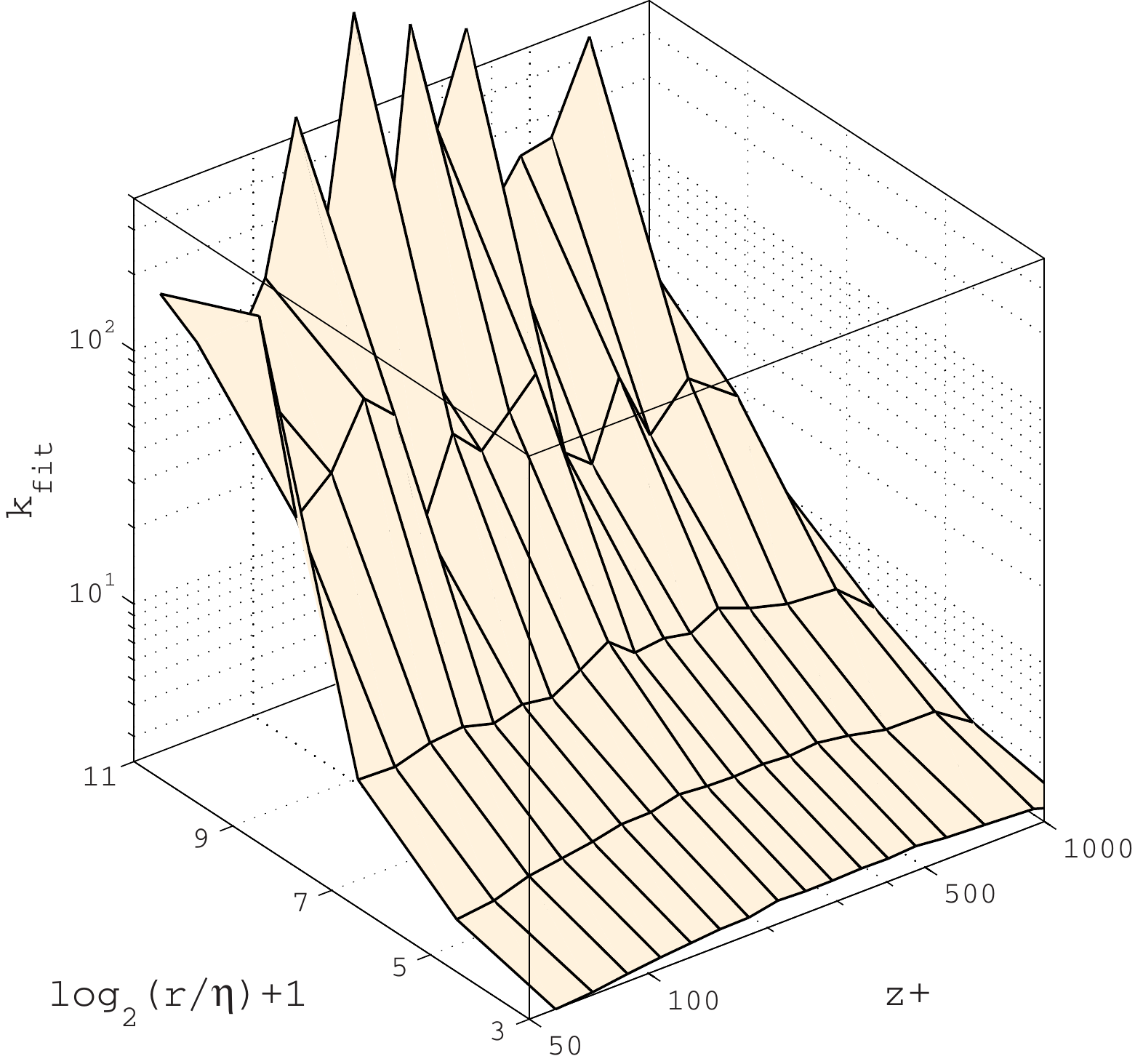}  
  \includegraphics[width=.49\textwidth]{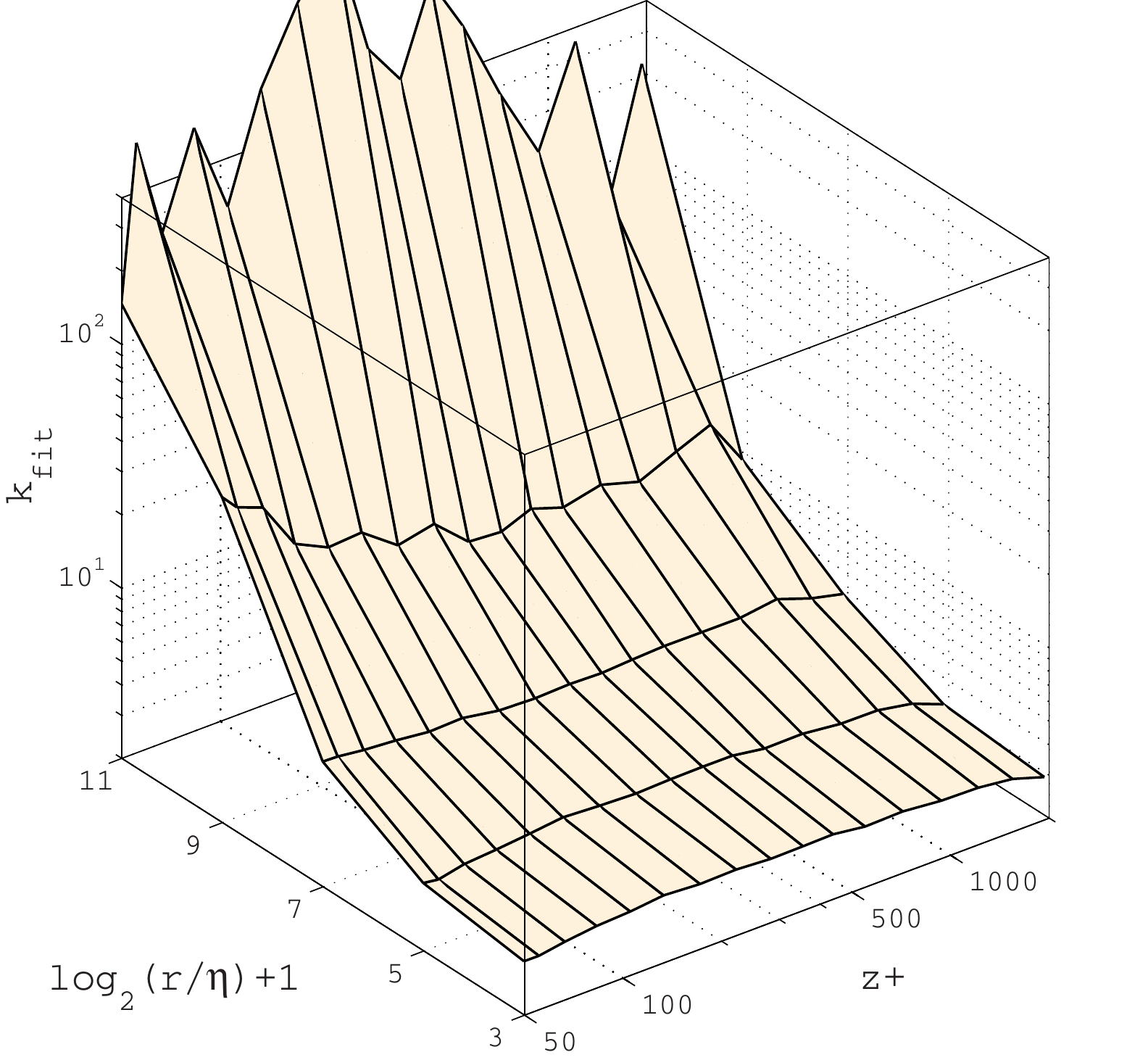} \\
  \includegraphics[width=.49\textwidth]{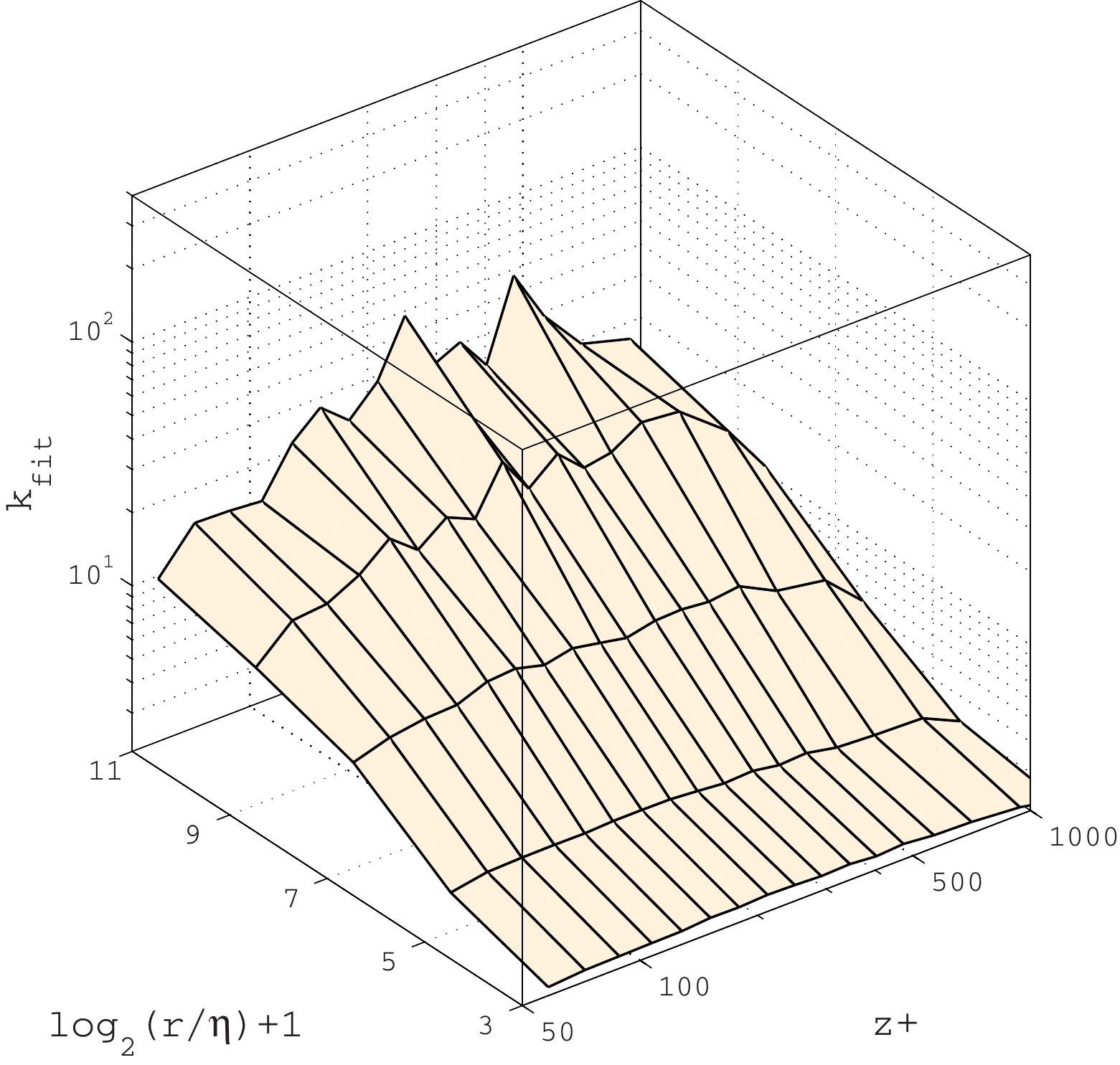}  
  \includegraphics[width=.49\textwidth]{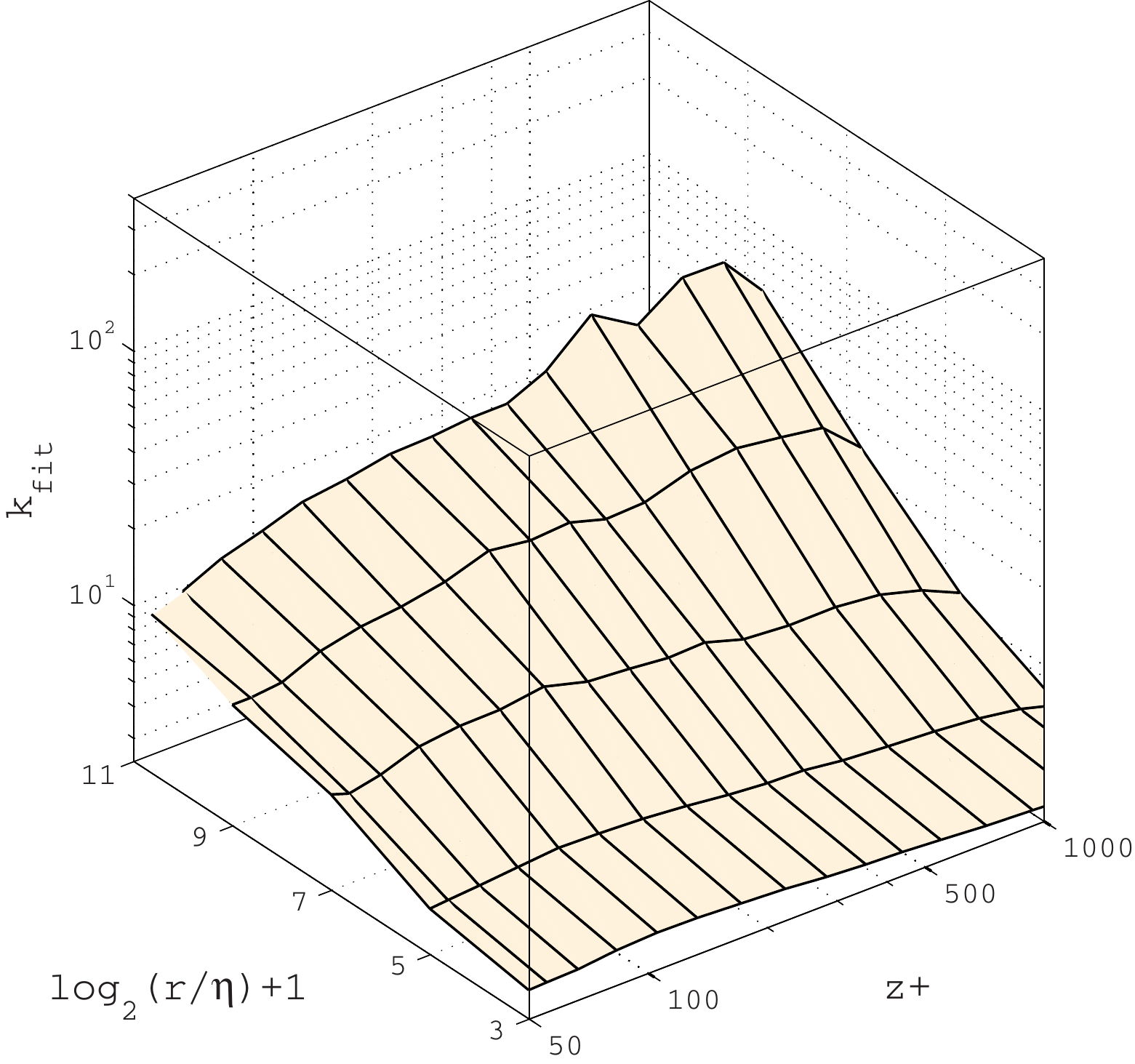}
  \caption{Best fits of $k$ for the streamwise (above) and wall-normal velocity component (below) at $\mathrm{Re}_\theta=4350$ (left) and $12600$ (right).}
  \label{fig:kfits}
\end{figure}

The data are not inconsistent with the suggested scaling $C\sim \mathrm{Re}_\lambda^{-1/2}$. The skewness observed is similar at both $\mathrm{Re}_\theta$, but $\mathrm{Re}_\lambda$ is also so similar between the two different data sets that this is no real test of this proposed scaling. We can only observe that the fitted values of $C$ for the streamwise components have the same order of magnitude as $\mathrm{Re}_\lambda^{-1/2}$ in the two cases considered. Taking the point $z^+ =105$ as example, $\mathrm{Re}_\lambda^{-1/2}=0.069$ here, while fitted values of $C$ lie in the range $-0.03$--$-0.056$. The same trend is found at $\mathrm{Re}_\theta=4350$: $C_\text{fit}\mathrm{Re}_\lambda^{1/2}\sim -0.6$ for $\log_2(r/\eta)+1 \lesssim 9$ seems to be the rough trend in the log-layer (though variations occur, as is in the nature of such imperfect data fits).

Examples of fits produced by fitting parameters $k$ and $C$ to experimental data by a least squares algorithm are shown at $z^+=105$ in figure \ref{fig:bestfits}. The corresponding values of $k$ and $C$ are tabulated in Table \ref{tab:kC}. Just as previously seen, the values of $k$ which produce best fits for the wall-normal velocity component roughly follow the $\log_2(r/\eta)+1$ trend, whereas the streamwise component converges to a near-Gaussian shape more quickly. As previously discussed, the cascade might only be presumed to be a reasonable cartoon picture of the real process whenever $\log_2(r/\eta)+1\lesssim 9$ or thereabouts, hence this behaviour is as expected. On the other hand it is surprising that this particular scaling appears to persist far beyond $\log_2(r/\eta)+1\sim 7$ for the wall-normal component. 

The skewness coefficient $C$ is consistently an order of magnitude larger, and of opposite sign, for $v_x$ compared to $v_z$. It is a trend throughout, for both data sets, that its best values first increase sligntly with increaseing $r$  (more so at $\mathrm{Re}_\theta=12600$ than at $4350$) before decreasing as $r$ approaches the integral scale. Best fits of $C$ for $v_x$ are plotted as a function of $r/\eta$ and $z^+$ in figure \ref{fig:Cfits}, where this trend is made obvious. 

A striking feature is the relative constancy of the best $C$ values with respect to $z^+$ throughout the log-layer, while we observe $z$ dependence as either bordering regime is approached. Particularly for the $\mathrm{Re}_\theta=4350$ dataset, which extends into both neighbouring regimes, we note how the best fits curve near either edge of the log-layer while staying relatively constant for $100\lesssim z^+\lesssim 1000$. The same constancy may be observed with reasonable accuracy in the fits of $k$, as shown in figure \ref{fig:kfits}. The dependence of $k$  upon $r$ is seen to be reasonably linear for the wall-normal component (here in log-log scale) while the same is not at all true in the streamwise direction for separations $\sim L_{xx}^{(x)}$.

\section{Conclusions and final remarks}

A model for fully developed turbulence based on Tsallis--generalised statistical mechanics has been tested for wall-bounded turbulence in the logarithmic sub-layer, in the embodiment delveloped, in particular, by Beck \cite{beck00}. The theory presents a model probability density function (PDF) for differences in velocities measured a distance $r$ apart. The simplest, parameter-free model uses an ``energy'' proportional to velocity differences squared with a semi-phenomenological skewness perturbation, and the non-extensivity parameter $q$ is interpreted so that $k=1/(q-1)$ is the bifurcation level in a hierarchy of eddy break-ups, inspired by the well documented turbulence cascade picture. 

The model was tested for two sets of wind tunnel data, at $\mathrm{Re}_\theta=12600$ and $4350$.
For the case of the wall-normal velocity component, this apparently simplistic model works strikingly well without further modifications. Indeed, for this component the scaling of $k$ with $r$ roughtly follows the cascade rule for $r$ up to several orders of magnitude beyond the corresponding integral scale, a surprising result. For the streamwise component the scaling conforms with the cascade picture only up to $r$ of the order of the integral scale as one would expect. 

We perform least square fits of $k$ and the (presumed small) skewness coefficient $C$ for a number of distances $z^+$ (in wall units) to the wall and values of $r$. Optimum values of both $C$ and $k$ remain roughly constant with respect to $z^+$ throughout the logarithmic sub-layer (but not beyond on either side). The skewness coefficient for streamwise velocities varies weakly with $r$, a slightly increasing function for small $r$ and decreasing as $r$ approaches the integral scale. Values of $C$ for the wall-normal component are consistently an order of mangintude smaller than for the streamwise component, and of opposite sign.

\section*{Acknowledgements}

We have benefited greatly from discussions with Professor Christian Beck during the work here presented.


\end{document}